# Focused Azimuthally Polarized Vector Beam and Spatial Magnetic Resolution below the Diffraction Limit

**Mehdi Veysi, Caner Guclu, and Filippo Capolino***

*Department of Electrical Engineering and Computer Science, University of California, Irvine, California 92697, USA*
*Corresponding author: f.capolino@uci.edu*

An azimuthally electric-polarized vector beam (APB), with a polarization vortex, has a salient feature that it contains a magnetic-dominant region within which electric field ideally has a null while longitudinal magnetic field is maximum. Fresnel diffraction theory and plane-wave spectral (PWS) calculations are applied to quantify field features of such a beam upon focusing through a lens. The diffraction-limited full width at half maximum (FWHM) of the beam's longitudinal magnetic field intensity profile and complementary FWHM (CFWHM) of the beam's annular-shaped total electric field intensity profile are examined at the lens's focal plane as a function of the lens's paraxial focal distance. Then, we place a subwavelength dense dielectric Mie scatterer in the minimum-waist plane of a self-standing converging APB and demonstrate for the first time that a very high resolution magnetic field at optical frequency is achieved with total magnetic field FWHM of $0.23\lambda$ (i.e., magnetic field spot area of $0.04\lambda^2$) within a magnetic-dominant region. The theory shown here is valuable for development of optical microscopy and spectroscopy systems based on magnetic dipolar transitions which are in general much weaker than their electric counterparts.

## I. INTRODUCTION

Vector beams [1]–[7] are a class of optical beams whose polarization profiles on the transverse plane, perpendicular to the beam axis, can be engineered to have an inhomogeneous distribution. Among them, beams with cylindrical symmetry (so-called cylindrical vector beams), particularly radially [3], [4], [8]–[10] and azimuthally [11]–[13] electric-polarized vector beams, are exceptionally important in the optics community. Owing to the presence of the longitudinal electric field component, a radially polarized vector beam with ring-shaped field profile after tight focusing through a lens provides a tighter electric field spot compared to the well-known linearly and circularly polarized beams [8], [9]. Such a beam has been extensively examined under tight focusing and has found many prominent applications in particle manipulation, high-resolution microscopy and spectroscopy systems [3], [5], [6], [8], [9], [14]–[24]. Here, we are particularly interested in studying the azimuthally electric-polarized vector beam primarily due to its unique magnetic field features, a strong longitudinal magnetic field where the electric field is null. In the following, we denominate such a beam simply as *azimuthally polarized beam (APB)* referring to the local orientation of its electric field vector. APBs possess an electric field purely transverse to the beam axis and a strong longitudinal magnetic field component in the vicinity of the beam axis where the transverse electric and magnetic fields are negligible and even vanish on the beam axis [12]. This so-called magnetic-dominant region is characterized by the presence of a tight magnetic field with longitudinal polarization. Especially, focusing an APB through a lens boosts its longitudinal magnetic field component relatively more than its transverse electric and magnetic fields [12]. Due to such unique property, the APB represents an intriguing choice for future spectroscopy and microscopy systems based on magnetic dipolar transitions [11], [12], [25]. Even though various methods have been proposed to generate APBs [12], [26]–[34], characterization of the magnetic field of these beams under tight focusing, to the authors' best knowledge, remains to be fully elucidated. This study is the basis for the successful implementation of magnetically sensitive nanoprobes at optical frequency which are crucial in the development of magnetism-based spectroscopy applications and the study of weak photoinduced magnetism in matter.

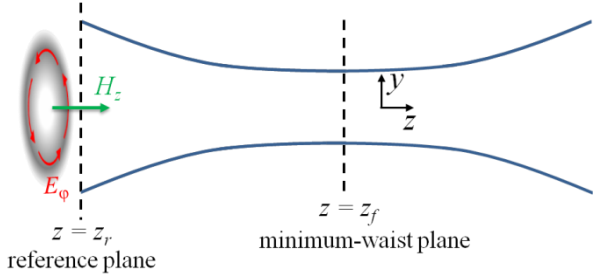

Fig. 1. Schematic of a converging azimuthally electric-polarized vector beam (APB), with a longitudinal magnetic field on its axis.

In this paper we report the diffraction-limited tight field (especially magnetic field) features of an APB, represented in terms of paraxial Laguerre Gaussian (LG) beams, with beam parameter $w_0$ that is a measure of the spatial extent of the beam in the transverse plane at its minimum waist. The two main figures of merit used in quantifying the field features in this paper are the full width at half maximum (FWHM) of the longitudinal magnetic field intensity and the complementary FWHM (CFWHM) of the annular-shaped total electric field intensity. Keeping in mind that for a very small beam parameter $w_0$ the expressions obtained via paraxial approximation may not be accurate, we also report results using the accurate analytical-numerical plane-wave spectral (PWS) calculations [35].

We first elaborate on the diffraction-limited tight focus of an APB through a converging lens using both paraxial Fresnel diffraction integral formulation, leading to analytical assessments, and the accurate PWS calculations (see [12] for more details on PWS). We demonstrate using the Fresnel integral under *paraxial approximation* that upon focusing through a lens an incident APB converts to another APB whose beam parameter is linearly proportional to the lens paraxial focal distance and inversely proportional to the incident APB parameter. The minimum-waist plane position of the focused beam predicted by the Fresnel integral coincides with the lens paraxial focal plane, which deviates from the actual focal plane position calculated by PWS. The figures of merit of an APB focused by a lens are therefore calculated both by the Fresnel integral at the lens's paraxial focus and by the PWS at the actual focal plane as a function of the lens paraxial focal distance.

In addition to the case of focusing an APB by a lens mentioned above, the tight field features of a self-standing converging APB are also examined and its figures of merit are calculated using the paraxial LG beam expressions and the PWS calculations at the minimum-waist planes predicted by the respective methods. Recently it has been experimentally confirmed that cylindrical vector beams may selectively excite the electric or magnetic dipolar resonances of a subwavelength-sized dense dielectric nanosphere (e.g., a silicon nanosphere) [24]. In this paper, we place a silicon nanosphere at the focus of a converging APB which selectively excites a magnetic dipolar resonance in the nanosphere as in [24] and focus on obtaining subwavelength magnetic field resolution. In general, such a subwavelength-sized scatterer hosts a magnetic Mie resonance with a circulating electric displacement current in addition to an electric dipolar resonance. However the latter is not excited by an APB due to its cylindrical symmetry, which ideally leads to a null average displacement current over the nanosphere. The induced electric displacement currents with a net magnetic dipole moment in the Si nanosphere along the z direction are shown to boost not only the total longitudinal magnetic field but also the spatial magnetic field resolution below the diffraction limit in the vicinity of the scatterer. The effective excited magnetic moment leads to a total magnetic field spot area as small as $0.04\lambda^2$ within a magnetic-dominant region, evaluated at a transverse plane one nanosphere radius ($0.12\lambda$) away from the scatter surface.

Throughout the paper we consider time harmonic fields with an $\exp(i\omega t)$ time dependence, which is suppressed for convenience. Furthermore bold symbols denote vectors and hats (^) indicate unit vectors.

## II. CHARACTERIZATION OF AN APB

APB is here expressed as a superposition of a left and a right hand circularly polarized beam, carrying orbital angular momentum (OAM) with orders of +1 and -1, respectively. In paraxial regimes, OAM-carrying beams are analytically represented as LG beams [1]. Thus the APB's electric field is expressed in terms of self-standing paraxial LG beams in cylindrical coordinate system as [12]

$$\mathbf{E} = \frac{-i\sqrt{2}}{2}\left(u_{-1,0}\hat{\mathbf{e}}_{RH} - u_{1,0}\hat{\mathbf{e}}_{LH}\right)e^{ikz}, \quad (1)$$

where $\hat{\mathbf{e}}_{RH} = (\hat{\mathbf{x}} + i\hat{\mathbf{y}})/\sqrt{2}$ and $\hat{\mathbf{e}}_{LH} = (\hat{\mathbf{x}} - i\hat{\mathbf{y}})/\sqrt{2}$ are, respectively, right and left hand circularly polarized unit vectors and the LG beam expression is

$$u_{\pm 1, p=0} = \frac{V}{\sqrt{\pi}} \frac{2\rho}{w^2} e^{-(\rho/w)^2 \zeta} e^{-2i\tan^{-1}(z/z_R)} e^{\pm i\varphi}, \quad (2)$$
$$w = w_0\sqrt{1+(z/z_R)^2}, \quad \zeta = (1 - iz/z_R)$$

where $V$ is an amplitude coefficient, $z_R = \pi w_0^2 / \lambda$ is the Rayleigh range, and $k = 2\pi/\lambda$ and $\lambda$ are the wavenumber and wavelength in the host medium, respectively. The beam parameter $w_0$ controls the transverse spatial extent of the beam at its minimum-waist plane. Vaguely speaking $w_0$ corresponds to the minimum waist which is very well defined for the fundamental Gaussian beam (FGB), but since the actual waist of the APB differs from $w_0$ we prefer to call it simply as "beam parameter" because this difference is of relevance in this paper. Here the term "beam waist" is reserved for the minimum of the actual waist size as discussed next.

The electric field in Eq. (1) is equivalently expressed as [12]

$$\mathbf{E} = E_\varphi \hat{\boldsymbol{\varphi}} = \frac{V}{\sqrt{\pi}} \frac{2\rho}{w^2} e^{-(\rho/w)^2} \zeta\, e^{-2i\tan^{-1}(z/z_R)} e^{ikz}\, \hat{\boldsymbol{\varphi}}, \quad (3)$$

that clearly shows the purely azimuthal polarization of the beam. Note that the choice of the sign in Eq. (3) is irrelevant. The electric field intensity profile of an APB is plotted at the beam's minimum-waist plane (i.e., $z = 0$) in Fig. 2(a). It is observed that the APB's electric field has an annular-shaped intensity profile whose CFWHM is of interest to us as a measure of the beam's tightness. The APB examined in Fig. (2) is carrying a power of $1\,\mathrm{mW}$, obtained by setting $V = 0.89\,\mathrm{V}$ in Eq. (3), and its beam parameter is set to $w_0 = 0.9\lambda$. The strength of the APB's electric field given in Eq. (3) is proportional to $\rho \exp(-\rho^2/w^2)$ in any given $z$ transverse plane, and it reaches its maximum

$$\left|E_\varphi(\rho_M, z)\right| = \frac{|V|}{\sqrt{\pi}} \frac{2\rho_M}{w^2} e^{-(\rho_M/w)^2} = \frac{\sqrt{2}}{\sqrt{\pi e}} \frac{|V|}{w}, \quad (4)$$

at $\rho_M = w/\sqrt{2}$. Therefore on the minimum-waist plane (i.e., $z = 0$) the electric field magnitude peaks at $\rho_M = w_0/\sqrt{2}$, that is in an agreement with what is shown in Fig. 2(a).

The magnetic field of the APB with the electric field given in Eq. (3) is subsequently found by using $i\omega\mu \mathbf{H} = \nabla \times \mathbf{E}$ in cylindrical coordinates, yielding a longitudinal magnetic field component as [12]

$$H_z = \frac{-V}{\sqrt{\pi}} \frac{4i}{w^2 \omega \mu} \left[1 - \left(\frac{\rho}{w}\right)^2 \zeta\right] e^{-\left(\frac{\rho}{w}\right)^2} \zeta\, e^{-2i\tan^{-1}\left(\frac{z}{z_R}\right)} e^{ikz}, \quad (5)$$

alongside a radial magnetic field component as

$$H_\rho = -\frac{1}{\eta} E_\varphi \left[1 + \frac{1}{kz_R} \frac{\rho^2 - 2w_0^2}{w^2}\right]. \quad (6)$$

It is observed from Eq. (6) that for $kz_R \gg 1$ the radial magnetic field component follows the electric field profile of the beam. In summary, the APB possesses only $E_\varphi, H_z$, and $H_\rho$ field components. The intensity of the APB's longitudinal magnetic field [given in Eq. (5)] is plotted in Fig. 2(b) where it peaks on the beam axis ($\rho = 0$) and is characterized by its FWHM. The maximum of the longitudinal magnetic field strength at any z is given by

$$\left|H_z(\rho = 0, z)\right| = 4|V|/\left(w^2 \omega \mu \sqrt{\pi}\right), \quad (7)$$

and is thus inversely proportional to $w^2$. It is observed from Eq. (7) that the longitudinal magnetic field of the APB peaks at the beam's minimum-waist plane (i.e., $z = 0$ where $w = w_0$), where its magnitude is inversely proportional to the square of the beam parameter $w_0^2$. The transverse magnetic field (which is purely radial and given in Eq. 6) increases together with the electric field (which is purely azimuthal) as the radial distance $\rho$ from the beam axis increases and peaks away from the beam axis alongside the azimuthal electric field, as shown in Fig. 2(c). By duality, this is analogous to the case of the radially polarized beam in which electric field intensity is purely longitudinal on the beam axis and its transverse component peaks off the beam axis [3], [4], [6], [11]. Here, we define the CFWHM for the annular-shaped electric field intensity profile of the APB as the width across its null where the field intensity rises to the half of its maximum, i.e., to $0.5\left|E_\varphi(\rho_M, z)\right|^2$ [see Eq. (4) and Fig. 2(a)]. In addition, the FWHM of the longitudinal magnetic field intensity is also calculated as the width across its peak on the beam axis where the longitudinal magnetic field intensity drops to the half of its maximum, i.e., to $0.5\left|H_z(\rho = 0, z)\right|^2$ [see Fig. 2(b)]. Based on the azimuthally polarized electric field and the longitudinally polarized magnetic field expressions given, respectively, in (3) and (5), the CFWHM of the electric field intensity and the FWHM of the longitudinal magnetic field intensity at the minimum-waist plane are calculated and given by

$$\begin{aligned}\mathrm{CFWHM}(E_\varphi)\big|_{z=0} &\approx 0.68 w_0,\\ \mathrm{FWHM}(H_z)\big|_{z=0} &\approx 0.81 w_0.\end{aligned} \quad (8)$$

One may be also interested in the ratio of the longitudinal magnetic field on the beam axis, where it is maximum, to the maximum of the electric field at $\rho = \rho_M$. This ratio, normalized with respect to the inverse of the host-medium wave impedance $\eta^{-1} = \sqrt{\varepsilon/\mu}$, is equal to

$$\eta \frac{\left|H_z(\rho = 0, z)\right|}{\left|E_\varphi(\rho_M, z)\right|} = \frac{\sqrt{2}}{\pi} \frac{\lambda}{w} e^{1/2} \approx 0.74 \frac{\lambda}{w}. \quad (9)$$

Note that such ratio is inversely proportional to $w$ and it reaches its maximum at $z = 0$, i.e., in the minimum-waist plane. Therefore the maximum magnitude of the longitudinal magnetic field increases relatively more than the maximum magnitude of the electric field as $w_0$ decreases (tighter beams). Note that decreasing $w_0$ also has the effect of decreasing the area of the longitudinal magnetic field spot.

Finally, we should note that on the minimum-waist plane ($z = 0$) the ratio $\eta\left|H_z(\rho, 0)\right|/\left|E_\varphi(\rho, 0)\right|$ is equal to unity at the radial distance

$$\rho = w_0 \left(\sqrt{1 + \left(\frac{\pi w_0}{2\lambda}\right)^2} - \frac{\pi w_0}{2\lambda}\right), \quad (10)$$

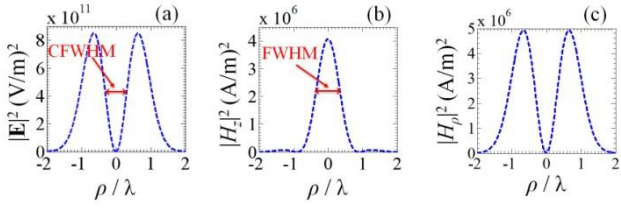

Fig. 2. Intensity profile of (a) electric field, (b) axis-confined longitudinal magnetic field, and (c) purely radial transverse magnetic field for an APB carrying $1\,\mathrm{mW}$ power and with beam parameter of $w_0 = 0.9\lambda$ at $\lambda = 523\,\mathrm{nm}$.

and inside this radius, the longitudinal magnetic to total electric field contrast ratio for APB is larger than the magnetic to electric field contrast ratio (the admittance) of a plane wave $1/\eta$. The optical power carried by the APB is calculated by the integral of its longitudinal Poynting vector over its minimum-waist plane as

$$P = \frac{1}{2}\int_0^{2\pi}\int_0^{\infty}\mathrm{Re}\left\{-E_\varphi\left(H_\rho\right)^*\right\}_{z=0}\rho\,d\rho\,d\varphi. \qquad (11)$$

After substituting the APB's azimuthal electric and radial magnetic fields formulas given in (3) and (5) into (11), the power carried by the APB is evaluated as

$$P = \frac{|V|^2}{2\eta}\left(1 - \frac{1}{(\pi w_0/\lambda)^2} + \frac{\int_0^{2\pi}\int_0^{\infty}\left(\frac{|E_\varphi|}{w_0|V|}\right)^2 \rho^3\,d\rho\,d\varphi}{2(\pi w_0/\lambda)^2}\right). \qquad (12)$$

After a change of variable from $2(\rho/w_0)^2$ to $t$, the integral term in Eq. (12) is found to be equal to $0.5\,\Gamma(3) = 1$ where $\Gamma(\cdot)$ is the gamma function. Therefore Eq. (12) is reduced to

$$P = \frac{|V|^2}{2\eta}\left(1 - \frac{1}{2(\pi w_0/\lambda)^2}\right). \qquad (13)$$

Eq. (13) clearly shows that power carried by the APB is explicitly expressed as a function of $w_0/\lambda$ and the absolute value of the amplitude coefficient $|V|$. Therefore, the APB's amplitude coefficient $V$ is obtained for certain beam parameter $w_0$ and required power using Eq. (13). The electric and magnetic field distributions of a paraxial APB and a FGB at their paraxial minimum-waist planes, i.e. at z = 0, are compared in Fig. 3 for two different beam parameters set to (top row) $w_0 = 0.9\lambda$ and (bottom row) $w_0 = 0.5\lambda$. Here the APB and the FGB carry equal powers of $1\,\mathrm{mW}$ [See Eq. (13)].

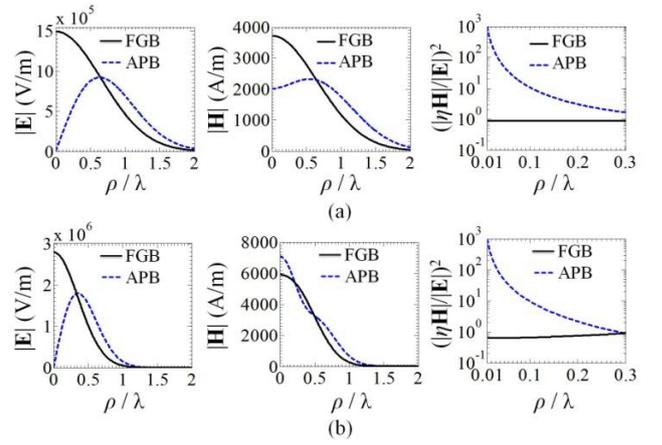

Fig. 3. Comparison between a FGB and an APB of equal powers (1mW) at $\lambda = 523\,\mathrm{nm}$ and beam parameters at their minimum-waist planes (i.e., $z = 0$): (top row) $w_0 = 0.9\lambda$ and (bottom row) $w_0 = 0.5\lambda$. Strength of the total electric field (first column), strength of the total magnetic field (second column), and the ratio of the total magnetic to the total electric field intensities normalized to that of a plane wave (third column) (Note how this ratio grows for the APB when approaching the beam axis.).

In order to have azimuthally symmetric magnetic field distribution for the FGB, we consider circularly polarized FGB in Fig. 3, however similar conclusions would be obtained if we used linearly polarized FGB. In contrast to the FGB whose magnetic and electric fields peak on the beam axis (i.e., the z axis), the APB contains a pure *longitudinal magnetic field component* on the beam axis where its electric field vanishes. The magnetic-to-electric field intensity ratio normalized to that of a plane wave is also plotted for the APB and the FGB in Fig. 3 (third column) varying radial distance from the beam axis. Note that the magnetic-to-electric field intensity ratio of the FGB is very close to that of a plane wave. In contrast the APB has a very large magnetic-to-electric field intensity ratio in the vicinity of the beam axis denoting the magnetic-dominant region. This ratio for the APB tends to infinity when $\rho \to 0$. For $w_0 = 0.9\lambda$ [Fig. 3(top row)], even though the strength of the APB's magnetic field on the beam axis is half of that of the FGB carrying the same power, the APB uniquely has only magnetic field and no electric field there, which is an important feature that can be used in various applications. In addition, it is observed from Fig. 3(top row) that the FWHM of the total magnetic field for the APB with $w_0 = 0.9\lambda$ is larger than that for the FGB with the same $w_0$, this is attributed to the fact that the APB contains an annular-shaped radial magnetic field component [see Fig. 2(c)]. However, decreasing the beam parameter (tightening the beam) from $w_0 = 0.9\lambda$ to $0.5\lambda$ boosts longitudinal magnetic field component relatively more than the radial one, and therefore significantly decreases the FWHM of the total magnetic field, as shown in Fig. 3(bottom row). To further reduce the FWHM of the total magnetic field of the APB approaching to that of its longitudinal magnetic field component, one approach might be to use ring-shaped lenses with high numerical apertures for focusing of the APB. This technique has been used for generating very sharp electric field focuses using radially polarized beams [8], [9].

### III. FOCUSING AN APB THROUGH A LENS

Let us now assume that an APB illuminates a converging lens. In Appendix A, using the Fresnel integral [Eq. (A3)] we show how a lens, under paraxial approximation, converts an incident APB, whose minimum-waist plane occurs at the lens surface, to another *converging* self-standing APB whose paraxial minimum-waist plane coincides with the lens's paraxial focal plane. This is schematically represented in Fig. 4 for a specific example where we show the total electric and the longitudinal magnetic field magnitudes of the APB before and after focusing through the lens.

In this section, the fields at the lens's paraxial focal plane are calculated by the paraxial Fresnel integral (See Appendix A for more details) and straightforwardly obtained as a function of the lens paraxial focal distance using Eq. (1) and (A10). To confirm the analytical calculations, we also characterize an APB upon focusing through a converging lens using the accurate PWS calculations. As for the PWS calculations, we assume the thin lens approximation such that each *ray* entering one side of the lens exits the other side at the same transverse $(\rho, \varphi)$ coordinates as the entrance position. We model the transmission through the lens by imposing a phase shift, which varies in radial direction, added to the $\rho$-dependent phase of the incident APB. The transmission phase shift that is added, relative to a spherically converging wave, is given by

$$\Phi(\rho) = -\frac{2\pi f}{\lambda}\left(\sqrt{1 + \frac{\rho^2}{f^2}} - 1\right), \quad (14)$$

where $f$ is the paraxial focal distance of the lens (on the right panel of Fig. 4), $\rho$ is the local radial coordinate of the lens, and $\lambda$ is the wavelength in the host medium on the right side of the lens. Note that we are assuming that the lens does not vary the $\rho-$dependent amplitude of the incident APB's field across the lens. We also remind that in the Fresnel integral equation the phase term in Eq. (14) is paraxially approximated as a quadratic phase [see Eq.(A2)] term [36].

We now characterize the FWHM of $|H_z|^2$ and the CFWHM of $|\mathbf{E}|^2$ at the lens focal plane for an incident APB. As a representative example we set the lens radius $a$ equal to $40\lambda$ and characterize the focusing beam at the lens's focal plane as the lens's paraxial focal distance $f$ changes. The *incident* APB has a beam parameter of $w_{0,i} = 29\lambda$ such that the beam cross-section is much wider than the wavelength and 90% of the incident beam power illuminates the lens surface. In Fig. 5 we plot the FWHM of $|H_z|^2$ and CFWHM of $|\mathbf{E}|^2$ calculated at the lens's focal plane as a function of the lens radius to focal distance ratio $a/f$, where $a$ is kept constant and $f$ is varied. We recall that the right side of Fig. 4 corresponds to the field maps for a specific case with $a = f/2$, which is a point on the curves reported in Fig. 5. The quantities plotted in Fig. 5 are calculated using both the Fresnel integral formula [given in Eq. (A10)] at the lens's paraxial focal plane ($z = f$) and PWS calculations (refer to [12] for more details on PWS) at the lens's actual focal plane.

It is observed from Fig. 5 that the paraxial Fresnel integral results (denoted by FI) agree very well with the accurate PWS results especially for large focal distances (small $a/f$). We also observe from the PWS results that for the case with $f = a$, the FWHM of the longitudinal magnetic field intensity and CFWHM of the total electric field intensity at the lens actual focal plane are $0.715\lambda$ and $0.53\lambda$, respectively. Note that the actual focal plane, obtained from PWS calculations, is slightly displaced from the lens's paraxial focal plane as described in the next section. In Appendix B, we elaborate more on this as we examine the plane wave spectrum of converging beams.

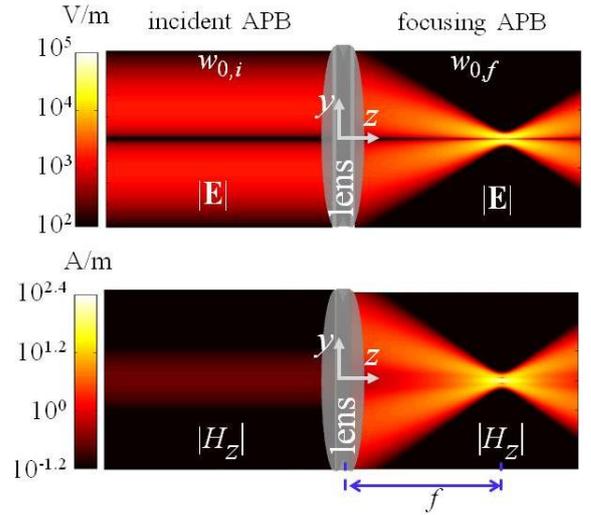

Fig. 4. Schematic of a converging lens transforming an incident APB with beam parameter $w_{0,i}$ into another converging APB with beam parameter $w_{0,f}$. The magnitudes of the total electric (which is purely azimuthal) and the longitudinal magnetic fields are plotted. The radial component of the magnetic field, also experiencing focusing, is not shown here for brevity. In this representative example, the incident APB carries $1\,\mathrm{mW}$ power, and the lens radius and focal distance are set at $a = 40\lambda$ and $f = 80\lambda$, respectively. The beam parameters of the incident and focusing APBs are $w_{0,i} = 29\lambda$ and $w_{0,f} = 1.3\lambda$, respectively.

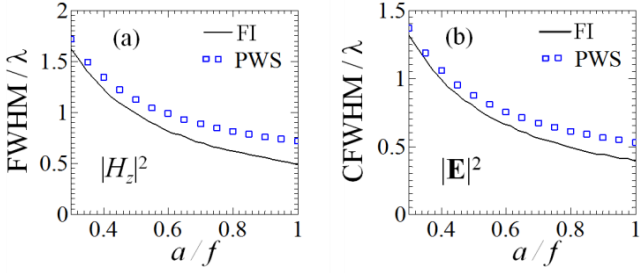

Fig. 5. (a) FWHM of the longitudinal magnetic field intensity $|H_z|^2$, and (b) CFWHM of the annular-shaped electric field intensity $|\mathbf{E}|^2$ calculated using (i) PWS at the actual focal plane and (ii) Fresnel integral (FI) at the lens's paraxial focal plane, upon illuminating the lens by an incident APB, varying normalized lens's focal distance $f$.

## IV. SELF-STANDING CONVERGING APB

As we discussed in the previous section, a lens converts an incident APB to another *converging* self-standing APB (see Fig. 4 and Appendix A). Calculations of the focusing beam due to a lens are simplified using the paraxial approximation (Appendix A). Therefore in this section we examine propagation of a self-standing converging APB, assuming it has been focused by the lens, and quantify its properties at its minimum-waist plane. We show some important tight field features of self-standing converging APBs as a function of their beam parameter $w_{0,f}$ (as shown in Fig. 4) where we pay particular attention to the FWHM of their longitudinal magnetic fields at the minimum waist. Since we only elaborate on a self-standing converging beam we drop the subscript $f$ and denote the converging beam parameter simply as $w_0$. The results pertaining to the paraxial beam propagation are compared to those obtained from the analytical-numerical computation based on the PWS. We assume to know the initial APB's field distribution, with converging features, at a certain *z*-plane (so-called reference plane) and observe the beam propagating toward its minimum-waist plane in +*z* direction. In other words, we investigate the converging properties of the beam on the right side of the lens in Fig. 4.

We first assess the validity of the paraxial approximation for APBs as in Eq. (1). It is known that the paraxial approximation for a beam holds under the following condition [2], [37], [38]

$$\left| 2k \frac{\partial \psi}{\partial z} \right| \gg \left| \frac{\partial^2 \psi}{\partial z^2} \right|, \quad (15)$$

where $\mathbf{E} = \boldsymbol{\psi} e^{ikz}$ represents paraxial field distribution for a beam propagating in +*z* direction. In order to determine the validity range of the paraxial field, we define a *paraxiality* figure as

$$F_p = \left| 2k \frac{\partial \psi}{\partial z} \right| \bigg/ \left| \frac{\partial^2 \psi}{\partial z^2} \right|, \quad (16)$$

which is a function of local coordinate. We also define the *normalized weighted average figure* of the paraxiality at each transverse *z*-plane as

$$F_{p,\text{ave}} = \frac{\int_{-\infty}^{\infty} \int_{-\infty}^{\infty} F_p |\psi|^2 \, dx \, dy}{\int_{-\infty}^{\infty} \int_{-\infty}^{\infty} |\psi|^2 \, dx \, dy}, \quad (17)$$

where the numerator is the average paraxiality figure weighted by the intensity of the transverse field and the denominator is the total weight of the transverse field intensity with respect to which we normalize the weighted average paraxiality figure. The value of $F_{p,\text{ave}}$ for the paraxial APB's electric field in Eq. (1) is calculated and plotted in Fig. 6 as a function of the beam parameter $w_0$, at the beam's paraxial minimum-waist plane $z = 0$. The larger the paraxiality figure $F_{p,\text{ave}}$ is, the better the paraxial approximation is. It is observed from Fig. 6 that $F_{p,\text{ave}} \geq 50$ (i.e., $\log_{10}(F_{p,\text{ave}}) \geq 1.7$) for the beam parameters larger than $0.9\lambda$. We assume that $F_{p,\text{ave}}$ values larger than 50 represent reasonably valid paraxial beams for practical purposes. Thus, for such values of $w_0$ the paraxial electric field expression given in Eq. (1) represents a self-standing APB's field distribution with a good approximation. Remarkably, the signature of this "validity range" manifests itself in the comparison of the paraxial beam propagation and the accurate PWS results discussed in the following.

We now examine the magnetic and electric field features of a self-standing converging APB at its minimum-waist plane as a function of the beam parameter $w_0$. With this in mind, we characterize self-standing converging APB using PWS calculations. We start with an APB's paraxial transverse field distribution on a transverse reference plane located at $z = z_r$ ($z_r < z_f$, see Fig. 1 for $z_f$) given by Eq. (1). Subsequently the evolutions of the beam's magnetic and electric fields in the positive $z$ direction are examined using the PWS calculations. The location of the actual minimum-waist plane of a converging APB "launched" from a reference plane at $z_r = -3.5\lambda$ with the field distribution given in Eq. (1) is calculated using PWS and plotted in Fig. 7 as a function of the beam parameter $w_0$. We observe that the actual minimum-waist plane of the converging APB does not occur at $z = 0$, that is the location of the focus predicted by the paraxial field expression. This difference is attributed to the presence of plane-wave constituents with large transverse wave numbers in the field spectrum of the converging APB which are not properly modeled in the paraxial field expressions (See Appendix B for more details on spectral content of the APB). For an APB, with decreasing $w_0$ a larger amount of constitutive propagating plane wave spectral components of the beam's field will have large transverse wavenumbers.

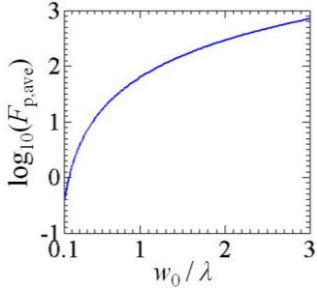

Fig. 6. The normalized weighted average figure of paraxiality $F_{p,ave}$ (in logarithmic scale) for a converging APB at the beam's paraxial minimum-waist plane ($z=0$) as a function of the beam parameter $w_0$.

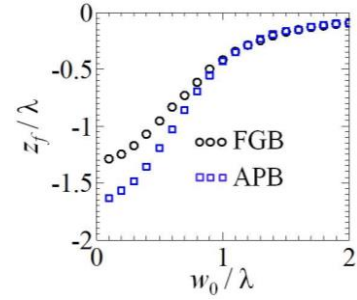

Fig. 7. The position of the actual minimum-waist plane of the beam $z_f$ as a function of the beam parameter $w_0$ for both APB and FGB (calculated using the PWS). The minimum-waist plane estimated by using the simple paraxial field expression is at $z=0$.

Hence, the difference between the actual minimum-waist plane's location (here denoted by $z_f$) and the one predicated by the paraxial field expressions (here at $z = 0$) becomes more significant, because of the loss of accuracy of the paraxial approximation with decreasing $w_0$. As a reference, in Fig. 7 we also plot the minimum-waist plane's position of a converging FGB as a function of its $w_0$. We observe that the difference between the actual and the paraxial minimum-waist plane's positions for a FGB is also increasing as $w_0$ decreases. However, the difference between the actual minimum-waist plane's position and the paraxial one is larger for an APB than for a FGB with an equal $w_0$. This is due to the fact that the transverse wavenumber spectrum of an APB's field is broader than that of a FGB's field with the same beam parameter; hence the paraxial approximation is coarser for the APB compared to the FGB.

Next the FWHM of $|H_z|^2$ and the CFWHM of $|\mathbf{E}|^2$ of the converging APB are calculated using both paraxial and PWS calculations and plotted in Fig. 8. The FWHM and the CFWHM in the PWS calculations are evaluated at the actual minimum-waist plane of the APB ($z = z_f$), that depends on $w_0$ (see Fig. 7). Instead, the FWHM and CFWHM under the paraxial approximation are evaluated using Eq. (1) at $z = 0$ for all the $w_0$ cases. It is observed from the paraxial calculations that the FWHM and CFWHM curves decrease monotonically as the beam parameter $w_0$ decreases. However, in practice, the decrease in FWHM of the longitudinal magnetic field intensity profile as well as CFWHM of the electric field intensity profile is hampered by an ultimate limit imposed by the diffraction of the beam. It is observed from the PWS curves in Fig. 8 that the FWHM and CFWHM of the converging APB are saturated by the diffraction to about $0.48\lambda$ and $0.37\lambda$, respectively, despite the paraxial approximation estimates much smaller FWHM and CFWHM. According to accurate PWS calculations, longitudinal magnetic field intensity profile with FWHM as small as $0.56\lambda$ (spot area of about $0.25\lambda^2$) is achievable with $w_0 = 0.5\lambda$. The spot area is defined here as the circular area whose diameter is equal to the FWHM.

Further decreasing $w_0$ from $0.5\lambda$ to $0.1\lambda$ only decreases the FWHM and CFWHM by about 15% considering the accurate PWS calculations. Here, based on what discussed in Appendix B, we stress that the transverse wavenumber spectrum of the APB's field in Eq. (1) with very small $w_0$ ($w_0 < 0.5\lambda$) is not confined only in the propagating wavenumber spectrum and it starts to extend to the evanescent spectral region.

The spatial field distribution in Eq. (1) with $w_0$ as small as $0.5\lambda$ has wavenumber spectral constituents still confined in the propagating spectrum (see Appendix B) and it has a relatively large normalized weighted average figure of paraxiality $F_{p,ave} \approx 14$. Therefore, though it may not represent a strictly self-standing APB, the paraxial approximation is not too coarse.

When the beam parameter $w_0$ is larger than $0.9\lambda$, the paraxial curves for the FWHM and the CFWHM in Fig. 8 follow very well the accurate PWS ones, and they start to deviate from PWS curves when $w_0$ decreases to smaller values, which is in agreement with our finding in Fig. 6. In order to clarify the effect of beam parameter $w_0$ on different magnetic field components of the APB, in Fig. 9 we plot the strength of the longitudinal ($H_z$) and the radial ($H_\rho$) magnetic field components as well as the strength of the azimuthal electric field normalized to the host-medium wave impedance for two different $w_0$ values at $\lambda = 523$ nm, using PWS calculations. We recall that the APB has a $H_z$ profile that peaks on the beam axis ($\rho = 0$), whereas its transverse magnetic field component is purely radial and peaks off the beam axis. It is observed from Fig. 9 that the longitudinal magnetic field spot areas as small as $0.23\lambda^2$ and $0.45\lambda^2$ are obtained with converging APBs with $w_0$ of $0.5\lambda$ and $0.9\lambda$, respectively. However, since APB possesses a radial magnetic field component over an annular-shaped region in addition to the longitudinal one (see Fig. 9), the FWHM of the total magnetic field is always larger than the FWHM of the longitudinal magnetic field component.

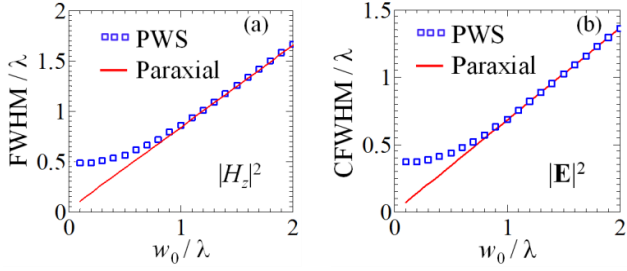
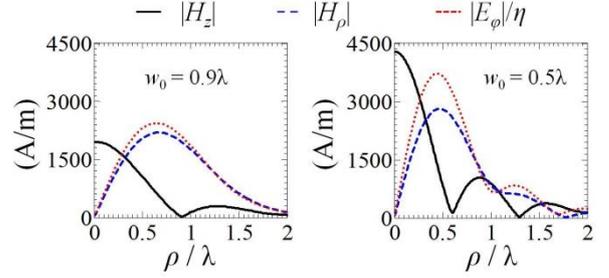

Fig. 8. Plane-wave spectral (PWS) and paraxial calculations for (a) the FWHM of the *longitudinal magnetic field* intensity and (b) the CFWHM of the annular-shaped electric field intensity of the converging APB as a function of the beam parameter $w_0$.

Fig. 9. Strength of the longitudinal ($H_z$) and radial ($H_\rho$) magnetic fields of an APB for two different beam parameters $w_0$ at $\lambda = 523\,\text{nm}$, evaluated using accurate PWS calculations. The strength of the azimuthal electric field ($E_\varphi$) normalized to the wave impedance is also plotted for comparison.

It is also observed from Fig. 9 that when $w_0$ of the APB decreases from $0.9\lambda$ to $0.5\lambda$, the strength of its longitudinal magnetic field component increases by about 2.2 times, which is relatively more than the increase in the strength of its radial magnetic field component (1.24 times). Indeed, as the beam parameter $w_0$ decreases, the plane-wave spectral distribution of the APB includes large transverse wavenumbers. For smaller beam parameters such like $w_0 = 0.5\lambda$, a larger portion of the constitutive TE (transverse electric with respect to z) plane waves in the spectrum of the APB possess large transverse wavenumbers meaning that they propagate in directions with larger angles $\alpha$ with respect to the beam axis, as shown in Fig. 10. Therefore the magnetic fields of the TE constitutive plane waves, which are perpendicular to the plane wave propagation directions, are more aligned with the beam axis. The depth of focus (DOF, or longitudinal FWHM) of the longitudinal magnetic field intensity profile for a converging APB is shown in Fig. 11 as a function of $w_0$, using accurate PWS calculations. For the sake of comparison, in Fig. 11 we also plot the DOF of the electric field intensity profile for a converging circularly polarized FGB. As $w_0$ increases, the Rayleigh range $z_R$ increases as $(w_0)^2$ and as a result the beam waist $w$ varies less with respect to z (see Eq. (2) where $w$ is written as a function of $z$ and $z_R$). Therefore the DOF is much longer for larger $w_0$ which also means the field features are less tight.

## V. SPATIAL MAGNETIC RESOLUTION BELOW THE DIFFRACTION LIMIT

So far we have demonstrated that the focusing features of a converging APB are limited by the diffraction limit. We have also shown that the total magnetic field intensity is less collimated than the longitudinal magnetic field due to the presence of the strong annular-shaped transverse magnetic field. In this section we aim at enhancing the longitudinal magnetic field of an APB and boosting its spatial magnetic resolution below the diffraction limit. To overcome the diffraction barrier, evanescent waves should be excited. One popular approach to generate evanescent waves, required for achieving spatial resolutions below the diffraction limit in microscopy, is to use a subwavelength scatterer [39].

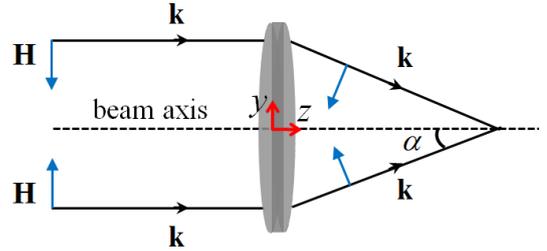

Fig. 10. Raytracing model of an APB focusing through a lens. Magnetic field vectors are denoted by blue arrows. Spectral components with large transverse wavenumber provide strong longitudinal magnetic field.

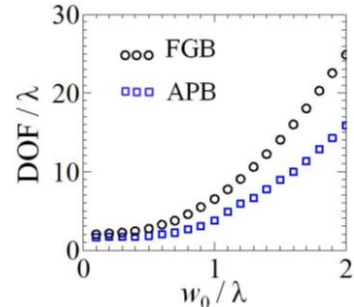

Fig. 11. The depth of focus (DOF) of the *longitudinal magnetic field* intensity profile for a converging APB as a function of the beam parameter $w_0$, evaluated using PWS. For comparison, the depth of focus of the electric field intensity profile of a FGB is also plotted.

Here, we show that a super tight magnetic-dominant spot is achieved using a subwavelength-size dense dielectric Mie (Silicon nanosphere) scatterer having a "magnetic" Mie resonance. We adopt an initial paraxial electric field distribution for the APB with $w_0 = 0.9\lambda$ at a reference plane $z = z_r$ (here $z_r = -3.5\lambda$) away from the minimum-waist plane based on Eq. (1), as shown in Fig. 12. In the previous section the propagation of such an APB was modeled using the PWS and its accurate minimum-waist plane position and field distributions are calculated. Here, we import the aforementioned APB's transverse electric field distribution into the finite integration technique in time domain solver implemented in CST Microwave Studio as a boundary field source. As a consequence of the Schelkunoff equivalence principle (PEC equivalent) implemented in CST Microwave Studio, the APB propagates towards the $+z$ direction in Fig. 12. The coefficient $V$ in Eq. (2) is set to $0.89\,\text{V}$ such that the total power of the incident APB given in Eq. (13) is $1\,\text{mW}$. In the full-wave simulations we assume a free space wavelength of $\lambda = 523\,\text{nm}$. The magnetic field map of the incident APB (without any scatterer yet), calculated by the time domain solver implemented in CST Microwave Studio in the *y-z* longitudinal plane, is shown in Fig. 13(a). We observe from Fig. 13(a) that the APB's minimum-waist plane occurs at $z_f \approx -0.55\lambda$ which is also obtained by the PWS calculations according to Fig. 7 (the paraxial approximation instead would estimate a focus at $z = 0$). Next, a subwavelength-size Silicon nanosphere scatterer is placed at the APB's actual minimum-waist plane ($z = z_f = -0.55\lambda$), assumed to be in vacuum. The Silicon nanosphere has a relative permittivity equal to $\varepsilon_r = 17.1 + i\,0.084$ and radius of $r = 62\,\text{nm}$, such that its magnetic Mie polarizability magnitude peaks at $\lambda = 523\,\text{nm}$ [25]. The total magnetic field's magnitude at the presence of nanosphere (superposition of incident and scattered fields) locally normalized to the magnitude of the incident magnetic field is also shown in the *y-z* longitudinal plane in Fig. 13(b), where we observe large magnetic field enhancement at the scatterer cross section and in the vicinity of the scatterer. Note that the enhanced magnetic field is strongly localized close to the scatterer resulting in a very high spatial magnetic resolution. Moreover, starting from the surface of the scatterer the tight magnetic spot extends into the surroundings and drops rapidly away from the nanosphere's surface, revealing the presence of evanescent spectral fields in the near field close to the scatterer.

The normalized magnetic and electric field intensities without the presence of the scatterer (only the incident APB) at the APB's minimum-waist plane (at $z = z_f$) and with the scatterer at two different transverse planes (at $z = z_f + r$ and $z = z_f + 2r$) slightly away from the scatterer are also shown in Fig. 14. High resolution magnetic field spots with FWHMs of $0.108\,\lambda$ and $0.233\,\lambda$ are obtained in the transverse planes tangential to the scatterer and one radius away from the scatterer in the positive *z* direction, respectively.

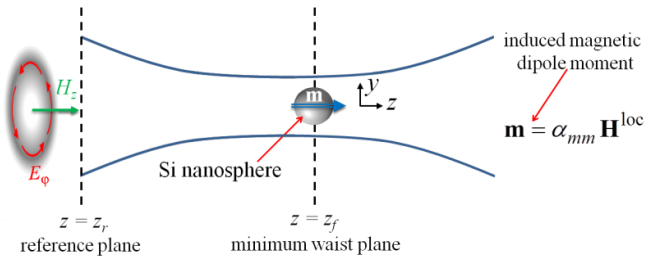

Fig. 12. Schematic of a converging APB with $w_0 = 0.9\lambda$ illuminating a subwavelength-size Silicon nanosphere placed at the actual minimum-waist plane of the beam: $r = 62\,\text{nm}$, $z_f = -0.55\lambda$, $\lambda = 523\,\text{nm}$.

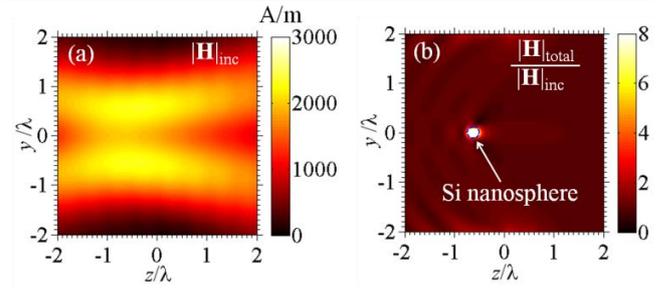

Fig. 13. Full-wave simulation results for the magnitude of (a) the incident magnetic field and (b) the total magnetic field (summation of incident and scattered field from the nanosphere) locally normalized to the incident magnetic field.

Such total magnetic field spot areas ($0.009\lambda^2$ and $0.043\lambda^2$) are much smaller than the ultimate spot area obtained for the longitudinal magnetic field of a tight focused APB without the nanosphere which is $0.43\lambda^2$. The enhancement of the total magnetic field with respect to that of the incident APB at two different transverse planes are also plotted in Fig. 15(a) where we observe a significant enhancement of the total magnetic field close to the nanosphere. The longitudinal magnetic field of the incident APB induces a magnetic dipole moment in the nanosphere polarized along the *z* direction which in turn boosts the total magnetic field thanks to the dipolar magnetic near fields. The square of the near-field admittance, defined as the total magnetic field intensity divided by the total electric field intensity, normalized to that of a plane wave ($1/\eta^2$) is also plotted in Fig. 15(b), which clearly shows a very high contrast ratio between magnetic and electric field especially around the beam axis. On the beam axis ($\rho = 0$) where the electric field has a null, magnetic to electric field contrast ratio goes to infinity (not shown here).

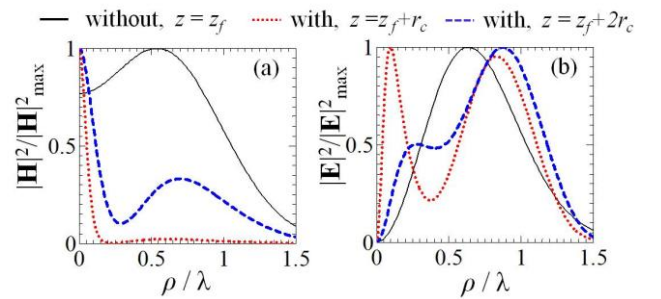

Fig. 14. Normalized total magnetic field intensity (each case is normalized to its own maximum) without (black solid curves) and with (blue dashed and red dotted curves) the presence of the Silicon nanosphere at $z = z_f$, evaluated at different transverse planes (*z* planes).

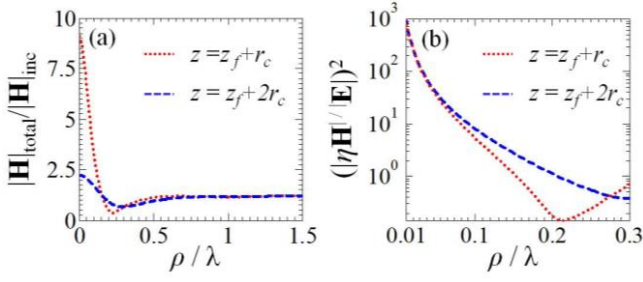

Fig. 15. (a) Total magnetic field (summation of incident and scattered fields) of the scatter system locally normalized to that without the nanosphere at $z = z_f$, evaluated at different transverse planes away from the scatterer. (b) Ratio of the total magnetic field intensity to the total electric field intensity of the scatter system normalized to that of plane wave (this defines the local near-field admittance normalized to that of the plane wave).

## VI. CONCLUSION

We have characterized the focusing of an azimuthally E-polarized vector beam (APB) through a lens with special attention on its magnetic-dominated region. When focusing the APB the longitudinal magnetic field strength grows relatively more than the azimuthal electric field strength, leading to a region of a boosted longitudinal magnetic field. We have also elaborated on self-standing converging APBs using plane-wave spectral (PWS) calculations and shown that longitudinal magnetic field intensity spot with full width at half maximum (FWHM) of $0.56\lambda$ and annular-shaped electric field intensity spot with complementary FWHM of $0.43\lambda$ can be achieved using a converging APB with $w_0 = 0.5\lambda$. However, the resolution of the total magnetic field intensity at a diffraction-limited APB focus is limited by the presence of the radial magnetic field in an annular-shaped region around the beam axis with comparable magnitude to the longitudinal one. In order to enhance the longitudinal magnetic field and obtain a very high total magnetic field resolution, we have proposed to utilize a magnetically polarizable (at optical frequency) particle leading to sharp magnetic near-field features. Full-wave simulation results reported here demonstrate that by placing a subwavelength-size dense Mie scatterer at the minimum-waist plane of a converging self-standing APB one achieves an extremely high resolution magnetic-dominant region with magnetic field spot area of $0.04\lambda^2$ at a transverse plane $0.12\lambda$ away from the scatterer surface. Such a super tight magnetic-dominant region with a negligible electric field is beneficial for future magnetism-based microscopy and spectroscopy systems.

## APPENDIX A: FIELD AT THE FOCAL PLANE OF A LENS UPON APB ILLUMINATION

Let us assume that an incident paraxial APB as in Eq. (1)-(2) with beam parameter $w_{0,i}$ illuminates an infinitely-thin converging lens (Fig. 4). We assume the lens to be positioned at $z = 0$ transverse plane where the incident APB has its minimum CFWHM. In other words, the lens is located at the incident beam's paraxial minimum-waist plane. Accordingly, the following conclusions would be still approximately valid if the incident beam's minimum-waist plane occurs at $|z| \ll z_R$ leading to $\zeta \approx 1$ and $w \approx w_0$ in Eq. (2). The electric field at the lens's paraxial focal plane $z = f$, on the right side of the lens in Fig. 4, is subsequently calculated using the Fresnel diffraction integral that in cylindrical coordinate system is written as [36]

$$\mathbf{E}_f(\rho,\phi,z=f) = \frac{k}{i2\pi f} e^{i\frac{k\rho^2}{2f}} e^{ikf} \int_0^\infty \int_0^{2\pi} P(\rho') \times$$
$$\times \left( \mathbf{E}_i(\rho',\phi',z'=0) e^{i\Phi(\rho')} e^{\left(i\frac{k\rho'^2}{2f}\right)} \right) e^{-i\frac{k}{f}\rho\rho'\cos(\phi-\phi')} \rho' d\rho' d\phi' \quad , \text{(A1)}$$

where $f$ is the lens paraxial focal distance, $\mathbf{E}_i(\rho',\phi',z'=0)$ is the incident APB's electric field vector at the lens plane (given by Eq. (1) with $z'=0$), $P(\rho')$ is the pupil function to account for the physical extent of the lens, and $\Phi(\rho')$ is the lens-induced spherical phase given in Eq. (14) required to focus the beam. Under paraxial approximation, the phase term $\Phi(\rho')$ in Eq. (14) is approximated as

$$\Phi(\rho') \approx -\frac{k\rho'^2}{2f} . \quad \text{(A2)}$$

Substituting Eq. (A2) into (A1), the electric field at the lens's paraxial focal plane can be subsequently approximated as

$$\mathbf{E}_f(\rho,\phi,z=f) \approx \frac{k}{i2\pi f} e^{i\frac{k\rho^2}{2f}} e^{ikf} \times$$
$$\times \int_0^\infty \int_0^{2\pi} P(\rho') \mathbf{E}_i(\rho',\phi',z'=0) e^{-i\frac{k}{f}\rho\rho'\cos(\phi-\phi')} \rho' d\rho' d\phi' \quad . \text{(A3)}$$

For simplicity, let us first assume that the physical diameter of the lens (i.e., $2a$) is sufficiently larger than the beam waist of the incident beam, implying that almost all the incident beam power illuminates the lens. Under such assumption, the pupil function in Eq. (A3) is set to one. The incident APB's electric field $\mathbf{E}_i(\rho',\phi',z'=0)$ in Eq. (A3) is a superposition of four linearly polarized LG beams, two $x$-pol. LG beams with $(l,p) = (\pm 1, 0)$ and two $y$-pol. LG beams with $(l,p) = (\pm 1, 0)$ [see Eq. (1)], where $l$ and $p$ are the azimuthal and radial LG beam's mode numbers, respectively. Therefore, for the sake of simplicity, we first show the steps for a general linearly polarized LG beam with mode number $(l,p)$ and beam parameter $w_{0,i}$ as the incident beam. Analogous treatment is readily applied to the all four linearly polarized LG beams that form the APB in Eq. (1). For an incident x-pol. LG beam, the electric field at the lens plane ($z=0$) is given as $\mathbf{E}_i(\rho',\phi',z'=0) = u_{l,p}(\rho',\phi',z'=0)\hat{\mathbf{x}}$. Here, we use the following integral identities [40]

$$\int_0^{2\pi} e^{il\phi'} e^{\left(\frac{-ik}{f}\rho\rho'\cos(\phi-\phi')\right)} d\phi' = 2\pi i^l e^{il\phi} J_l\left(\frac{k}{f}\rho\rho'\right)$$

$$\int_0^\infty \rho'^{\left(|l|+\frac{1}{2}\right)} e^{-\beta\rho'^2} L_p^{|l|}(\alpha\rho'^2) J_l\left(\frac{k\rho'\rho}{f}\right) \left(\frac{k\rho'\rho}{f}\right)^{\frac{1}{2}} d\rho' =$$

$$= [\text{sgn}(l)]^{|l|} 2^{-(|l|+1)} \beta^{-(|l|+p+1)} (\beta-\alpha)^p \left(\frac{k\rho}{f}\right)^{\left(|l|+\frac{1}{2}\right)} \times$$

$$\times e^{\left(-\frac{k^2\rho^2}{4\beta f^2}\right)} L_p^{|l|}\left[\frac{\alpha k^2 \rho^2}{4\beta f^2(\alpha-\beta)}\right]$$
(A4)

where $J_l(.)$ is the Bessel function of the first kind and order of $l$, $\text{sgn}(l)$ denotes the sign of $l$, and $L_p^{|l|}(.)$ is the associated Laguerre polynomial [1], [12]. Accordingly, the x-pol. LG beam's field at the paraxial focal plane of the lens is then written as

$$\mathbf{E}_f(\rho,\phi,z=f) \approx V[\text{sgn}(l)]^{|l|}(-1)^p \sqrt{\frac{2p!}{\pi(p+|l|)!}} e^{i\frac{k\rho^2}{2f}} \times$$

$$\times e^{ikf} \frac{i^{(l-1)}}{w_{0,f}} \left(\frac{\rho\sqrt{2}}{w_{0,f}}\right)^{|l|} e^{-\left(\frac{\rho}{w_{0,f}}\right)^2} L_p^{|l|}\left(\left(\frac{\rho\sqrt{2}}{w_{0,f}}\right)^2\right) e^{il\phi} \hat{\mathbf{x}}$$
(A5)

where

$$w_{0,f} = \frac{\lambda}{\pi} \frac{f}{w_{0,i}}.$$
(A6)

For practical cases (as for the lens physical parameters provided in this paper), almost all the focused beam power on the lens focal plane is confined in a circular region whose area is much smaller than $\pi f\lambda$, i.e., the focused area is within a radial distance $\rho \ll \sqrt{\lambda f}$. Therefore, the phase factor $\exp[ik\rho^2/(2f)]$ in Eq. (A5) can be neglected and Eq. (A5) would clearly represent a paraxial LG beam at its minimum-waist plane with beam parameter of $w_{0,f}$. In other words, under paraxial assumption, focusing an incident LG beam through a simple lens placed at its minimum-waist plane results in another LG beam whose minimum-waist plane coincides with the lens's focal plane ($z=f$) and its beam parameter relates to the focal distance of the lens and the incident beam parameter through the Eq. (A6). Note that, in principle and according to Eq. (A6), if the radial spread of the incident LG beam determined through $w_{0,i}$ is in comparable length to the focus distance $f$, then the beam parameter of the converged beam $w_{0,f}$ would be subwavelength. However as shown in Sec. IV, physical limitations accounted by using the PWS (see [12] for details on the PWS calculations) shows that there is a limit in the minimum achievable $w_{0,f}$.

The electric field at the lens paraxial focal plane due to an incident APB is subsequently obtained by the superposition of the focal plane fields of its four constitutive linearly polarized LG beam terms leading to

$$\mathbf{E}_f = \frac{V}{\sqrt{\pi}} \frac{2\rho}{w_{0,f}^2} e^{-\left(\frac{\rho}{w_{0,f}}\right)^2} e^{ikf} e^{i\frac{k\rho^2}{2f}} \hat{\boldsymbol{\varphi}}.$$
(A7)

In order to take into account for the physical extent of the lens, we also consider the following pupil function

$$P(\rho') = \begin{cases} 1 & \rho' \leq a \\ 0 & \rho' > a \end{cases},$$
(A8)

where $a$ is the lens radius. The pupil function in (A8) is expanded into a summation of Gaussian functions that come handy in taking the Fresnel integral in (A3) analytically. Such a pupil function is approximated with a finite summation of basis Gaussian functions as [41]

$$P(\rho') \approx \sum_{n=1}^{N} A_n e^{-\frac{B_n}{a^2}\rho'^2},$$
(A9)

where complex coefficients $A_n$ and $B_n$ are, respectively, expansion and Gaussian coefficients. It is demonstrated in [41] that for N = 10 the pupil function in (A8) is well represented by Eq. (A9) with proper coefficients given in [41]. By substituting (A9) into (A3), the focusing field at the focal plane $z=f$ of the finite-size lens upon illumination by an incident x-pol. LG beam with $(l,p)$ is calculated as

$$\mathbf{E}_f \approx V[\text{sgn}(l)]^{|l|} \sqrt{\frac{2p!}{\pi(p+|l|)!}} e^{i\frac{k\rho^2}{2f}} e^{ikf} \frac{i^{l-1}}{w_{0,f}} \left(\frac{\rho\sqrt{2}}{w_{0,f}}\right)^{|l|} e^{il\phi} \times$$

$$\times \sum_{n=1}^{N} A_n (1+\beta_n)^{-(|l|+p+1)} (\beta_n - 1)^p e^{-\frac{(\rho/w_{0,f})^2}{1+\beta_n}} L_p^{|l|}\left(\frac{\left(\frac{\rho\sqrt{2}}{w_{0,f}}\right)^2}{1-\beta_n^2}\right) \hat{\mathbf{x}},$$
(A10)

where $\beta_n = B_n w_0^2/a^2$, $w_{0,f}$ is given in (A6), and the summation over the Gaussian expansion index $n$ appears in the focal field distribution term. In this way the paraxial approximation of the focusing field at the lens focal plane due to an incident x-pol. LG beam is conveniently expressed in series terms (A10) for the case of a pupil function of finite extent. The electric field at the lens paraxial focal plane due to an incident APB illumination is then calculated by the superposition of the four constitutive linearly polarized LG beam terms.

## APPENDIX B: SPECTRAL INTERPRETATION OF THE BEAM PROPAGATION IN NON-PARAXIAL REGIME

The electric field distribution for APB given in Eq. (1) represents a self-standing beam in the paraxial regime. Therefore it is important to address limitations of these paraxial expressions in the cases of beams with very tight spatial extents (small $w_0$). With this goal in mind, here we report in Fig. 16 the normalized magnitude of the plane-wave spectrum for APBs, i.e., the 2-D Fourier transform of the transverse field of the APB as

$$\tilde{\mathbf{E}}(k_x, k_y, z) = \int_{-\infty}^{+\infty}\int_{-\infty}^{+\infty} \mathbf{E}(x,y,z) e^{-ik_x x - ik_y y} dx dy, \quad (B1)$$

(see [12] for details on the numerical calculation of the integral). In Fig. 16 we show the wavenumber spectrum of three APBs with different beam parameters: (a) $w_0 = 3\lambda$, (b) $w_0 = 0.9\lambda$, and (c) $w_0 = 0.5\lambda$. It is observed from Fig. 16 that the plane-wave spectrum of the tighter beam (beam with smaller $w_0$) covers a wider region in the $k_x - k_y$ plane where $k_0 = 2\pi/\lambda$ is the free space wavenumber. Moreover the field spectral distribution for all three beams is well confined in the propagating wave spectrum with $k_x^2 + k_y^2 < k_0^2$, hence they are mainly constructed by propagating spectral components only. They propagate along the z axis with $\exp(ik_z z)$ where $k_z$ is real and is evaluated as

$$k_z = \sqrt{k_0^2 - (k_x^2 + k_y^2)}. \quad (B2)$$

All spectral magnitude distributions in Fig. 16 are representative at any z plane as these field spectral distributions propagate with no magnitude variation (implied by the propagator with magnitude $|\exp(ik_z z)| = 1$).

Let us now consider these three field distributions and look at the paraxial wave approximation. The paraxial wave equation is valid under the assumption that most of the field spectrum is confined to a region with $k_x^2 + k_y^2 \ll k_0^2$. Under this condition the accurate PWS evaluation can be approximated with the paraxial field expression of a propagating beam as in Eqs. (1)-(2) [37]. Indeed, the required condition for deriving the paraxial field expressions using PWS calculations is to approximate Eq. (B2) as

$$k_z \approx k_0 - (k_x^2 + k_y^2)/(2k_0). \quad (B3)$$

It is observed from Fig. 16 that for $w_0 = 0.9\lambda$ and $w_0 = 0.5\lambda$ cases the spectral distributions cannot be fully confined to a region with $k_x^2 + k_y^2 \ll k_0^2$ in contrast to the case with $w_0 = 3\lambda$. Therefore, the prediction of the paraxial beam propagation is expected to deviate from the actual propagation of the beam, much more for $w_0 = 0.5\lambda$ than for $w_0 = 0.9\lambda$ and much more for $w_0 = 0.9\lambda$ than for $w_0 = 3\lambda$. The spectrums with $w_0 = 0.5\lambda$ and $w_0 = 0.9\lambda$ generate tight field spots but the z location of the tight spots cannot be accurately predicted by the paraxial field equations.

Especially, when considering a converging beam with $w_0 = 0.5\lambda$ or $0.9\lambda$, we can expect that the actual tight spot location (minimum-waist plane) will be formed closer to the reference plane than the one predicted by the paraxial expressions. This is due to the fact that the field of the APB with $w_0 = 0.5\lambda$ or $0.9\lambda$ constitutes plane-wave spectral components with large transverse wave numbers which are not modeled accurately in the paraxial field expressions and propagate at larger incidence angles with respect to the beam axis (the z axis) and thus a tight spot forms closer than the one predicted by paraxial field expressions. The time-average spectral energy of the APB per unit length along the z-direction is calculated as

$$\tilde{W} = \frac{1}{4}\varepsilon_0 |\tilde{\mathbf{E}}|^2 + \frac{1}{4}\mu_0 |\tilde{\mathbf{H}}|^2. \quad (B4)$$

We define here the figure of APB's spectral energy as the ratio of the APB's spectral energy per unit length in the propagating spectrum to its total spectral energy per unit length

$$F_W = \frac{\iint\limits_{k_x^2+k_y^2<k_0^2} \tilde{W} dk_x dk_y}{\int_{-\infty}^{+\infty}\int_{-\infty}^{+\infty} \tilde{W} dk_x dk_y}. \quad (B5)$$

Fig. 17 shows the figure of APB's spectral energy as a function of the beam parameter. It is observed that for $w_0 \geq 0.5\lambda$ more than 95% of the APB's spectral energy is confined in the propagating spectrum.

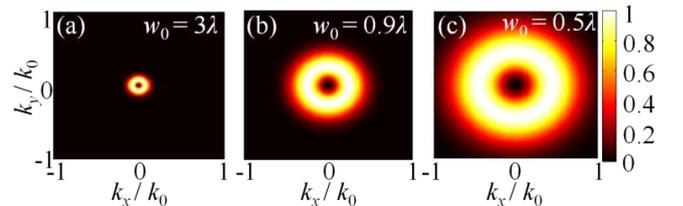

Fig. 16. Normalized magnitude of the transverse field spectrum $\tilde{\mathbf{E}}$ for APBs with (a) $w_0 = 3\lambda$, (b) $w_0 = 0.9\lambda$, and (c) $w_0 = 0.5\lambda$. Note that these APBs are made mainly by propagating spectrum (such that $k_x^2 + k_y^2 < k_0^2$), and therefore the spectral magnitude profiles are basically similar at any transverse plane (here we show only the propagating spectrum region).

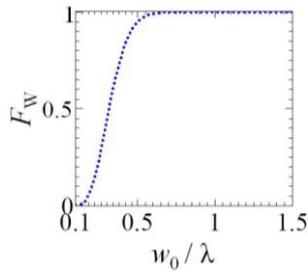

Fig. 17. Ratio of the APB's spectral energy per unit length (in z-direction) confined in the propagating spectrum to its total spectral energy per unit length (so-called figure of APB's spectral energy) defined in Eq. (B5) as a function of the beam parameter $w_0$.

## ACKNOWLEDGMENTS

The authors acknowledge support by the W. M. Keck Foundation. The authors would like to thank also Computer Simulation Technology (CST) of America, Inc. for providing CST Microwave Studio that was instrumental in this work.